\documentclass[]{iopart}
\usepackage[]{graphicx}
\begin{document}

\title[Quantum locking of mirrors]
{Beating quantum limits in interferometers with quantum locking of mirrors}

\author{Antoine Heidmann, Jean-Michel Courty, Michel Pinard and Julien Lebars}

\address{Laboratoire Kastler Brossel
\footnote{Unit\'{e} mixte de recherche du Centre National de la
Recherche Scientifique, de l'Ecole Normale Sup\'{e}rieure et de
l'Universit\'{e} Pierre et Marie Curie; URL:
www.spectro.jussieu.fr/Mesure}, Case 74, 4 place Jussieu, F75252
Paris Cedex 05, France}

\ead{heidmann@spectro.jussieu.fr}

\begin{abstract}
The sensitivity in interferometric measurements such as gravitational-wave
detectors is ultimately limited by quantum noise of light. We discuss the use
of feedback mechanisms to reduce the quantum effects of radiation pressure.
Recent experiments have shown that it is possible to reduce the thermal
motion of a mirror by cold damping. The mirror motion is measured with an
optomechanical sensor based on a high-finesse cavity, and reduced by a
feedback loop. We show that this technique can be extended to lock the mirror
at the quantum level. In gravitational-waves interferometers with Fabry-Perot
cavities in each arms, it is even possible to use a single feedback mechanism
to lock one cavity mirror on the other. This quantum locking greatly improves
the sensitivity of the interferometric measurement. It is furthermore
insensitive to imperfections such as losses in the interferometer.
\end{abstract}


\section{Introduction}

Quantum fluctuations of light play an important role in the sensitivity
limits of optical measurements such as interferometric measurements
considered for gravitational-wave detection \cite{Bradaschia90,Abramovici92}.
A gravitational wave induces a differential variation of the optical pathes
in the two arms of a Michelson interferometer. The detection of the phase
difference between the two optical pathes is ultimately limited by the
quantum noises of light: the phase fluctuations of the incident laser beam
introduce noise in the measurement whereas radiation pressure of light
induces unwanted mirrors displacements. Due to the Heisenberg's inequality,
both noises are conjugate and lead to the so-called standard quantum limit
for the sensitivity of the measurement when coherent states of light are used
\cite{Caves81,Jaekel90,Braginsky92}.

Potential applications of squeezed states to overcome this limit have
motivated a large number of works in quantum optics. The injection of a
squeezed state in the unused port of the interferometer can improve the
sensitivity of the measurement \cite{Jaekel90,Braginsky92,McKenzie02}.
Another possibility is to take advantage of the quantum effects of radiation
pressure in the interferometer to perform a quantum nondemolition measurement
\cite{Braginsky92}. Since radiation pressure effects are frequency dependent,
a main issue is to find simple schemes which improve the sensitivity over a
wide frequency band \cite{Kimble02}. Another issue is to precisely examine
the constraints imposed to the interferometer by the use of such quantum
techniques. Losses in particular may have drastic effects on the sensitivity
improvement.

An alternative approach consists in using feedback mechanisms working in the
quantum regime. This technique has been proposed to generate squeezed states
\cite{Haus86} or to perform quantum nondemolition measurements
\cite{Wiseman95}, and experimentally demonstrated on laser oscillators
\cite{Yamamoto86} and twin beams \cite{Mertz90}. Active controls are also
widely used in the classical regime as for example in cold-damped mechanical
systems \cite{Milatz53,Grassia00}. Cold damping is able to reduce the
mechanical thermal displacements of a mirror \cite{Cohadon99,Pinard00}, and
it may in principle be used to reduce the displacement noises in a quantum
regime, down to the zero-point quantum fluctuations of the mirror
\cite{Mancini00,Courty01}.

We discuss in this paper the possibility to increase the sensitivity in
interferometers by reducing radiation pressure effects with such a feedback
mechanism. We propose to use a compact optomechanical sensor made of a
high-finesse cavity to measure the mirror displacements induced by radiation
pressure. The information is fed back to the mirror in order to lock its
position at the quantum level. We show that the sensor sensitivity is
transferred to the interferometric measurement, resulting in a reduction of
the back-action noise associated with the radiation pressure in the
interferometer \cite{Courty03a}.

We present in section \ref{Sec_QuantumLimits} the main characteristics of the
quantum noises in an interferometric measurement. Section
\ref{Sec_QuantumLocking} is devoted to the active control of a mirror and to
the resulting reduction of back-action noise in the interferometer. As shown
in section \ref{Sec_BackAction}, it is possible to completely suppress
back-action noise if the optomechanical sensor performs a quantum
nondemolition measurement of the mirror motion \cite{Courty03b}. We finally
study in section \ref{Sec_CavityLocking} the control of the whole Fabry-Perot
cavity in each interferometer arms with a single optomechanical sensor. We
show that the information provided by the measurement allows one to lock the
cavity length in such a way that it is no longer sensitive to
radiation-pressure effects.

\section{Quantum limits in interferometric measurements}
\label{Sec_QuantumLimits}

\begin{figure}
\begin{center}
\includegraphics[width=7cm]{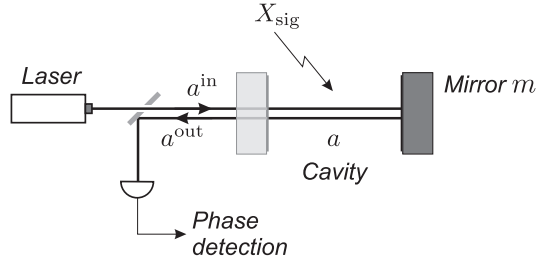}
\end{center}
\caption {\label{Fig_Cavity} An interferometric measurement is equivalent to
a length measurement by a single Fabry-Perot cavity. A cavity length
variation $X_{\rm sig}$ is detected through the phase shift induced on the
field $a^{\rm out}$ reflected by the cavity.}
\end{figure}

A gravitational-wave interferometer is based on a Michelson interferometer
with kilometric arms and Fabry-Perot cavities inserted in each arm
\cite{Bradaschia90,Abramovici92}. A gravitational wave induces a length
variation of the cavities and is detected as a change in the interference
fringes at the output of the interferometer. As long as we are concerned with
quantum and mirror-induced noises, the interferometer is equivalent to a
simpler scheme consisting in a single resonant optical cavity, as shown in
figure \ref{Fig_Cavity}, which actually corresponds to one interferometer
arm. A variation of the cavity length changes the optical path followed by
the intracavity field and induces a phase shift of the reflected field which
can be detected by an homodyne detection.

To study the effects of quantum fluctuations of light, we describe
the fields by a complex mean amplitude $\alpha $ and quantum
annihilation operators $a\left[ \Omega \right]$ at frequency
$\Omega $. We define any quadrature $a_\theta $ of the field as
\begin{equation}
a_\theta \left[ \Omega \right] =e^{-i\theta }a\left[ \Omega
\right] +e^{i\theta }a^{\dagger }\left[ \Omega \right] .
\label{Equ_Quad}
\end{equation}
For a lossless single-ended cavity resonant with the laser field,
the input-output relations for the fields can be written in a
simple way. The reflected mean field $\alpha ^{\rm out}$ is equal
to the incident mean field $\alpha ^{\rm in}$ and both can be
taken as real. The amplitude and phase quadratures of the fields
then correspond to the quadratures $a_0$ and $a_\frac{\pi }{2}$,
respectively aligned and orthogonal to the mean field. Assuming
the frequency $\Omega $ of interest smaller than the cavity
bandwidth, the input-output relations for the field fluctuations
are given by \cite{Courty01},
\begin{eqnarray}
a_0^{\rm out}=a_0^{\rm in}, \label{Equ_Ampl} \\
a_\frac{\pi }{2}^{\rm out}=a_\frac{\pi }{2}^{\rm in}+2\xi _{\rm
a}X, \label{Equ_Phase}
\end{eqnarray}
where $X$ is the cavity length variation and $\xi _{\rm a}$ is an
optomechanical parameter related to the intracavity mean field $\alpha $, the
cavity finesse ${\cal F}_{\rm a}$, and the optical wavelength $\lambda $,
\begin{equation}
\xi _{\rm a}=\frac{4\pi }{\lambda }\alpha \sqrt{2{\cal F}_{\rm
a}/\pi }. \label{Equ_Xia}
\end{equation}
Equations (\ref{Equ_Ampl}) and (\ref{Equ_Phase}) show that the reflected
fluctuations reproduce the incident ones, but the phase quadrature is also
sensitive to the cavity length variation $X$. Neglecting any mirror
displacement, this variation corresponds to the length change $X_{\rm sig}$
due to the gravitational wave. The sensitivity of the measurement is only
limited by the incident phase noise $a_\frac{\pi }{2}^{\rm in}$. For a
coherent incident field, quantum fluctuations are characterized by a noise
spectrum equal to $1$ for any quadrature. One thus expects to be able to
detect length variations small compared to the optical wavelength $\lambda $
by using large values of the optomechanical parameter $\xi _{\rm a}$, that is
for a high-finesse cavity and an intense incident beam.

Mirror displacements also limit the sensitivity of the measurement. In the
following we focus on the displacements of a single mirror of the cavity,
namely, the end mirror $m$ in figure \ref{Fig_Cavity}. The cavity length
variation $X$ in (\ref{Equ_Phase}) is then the sum of the signal $X_{\rm
sig}$ and the displacement $X_{\rm m}$ of mirror $m$, which corresponds to
the back-action noise due to radiation pressure, and to classical noises such
as seismic or thermal fluctuations. Its Fourier component at frequency
$\Omega $ is related to the applied forces by \cite{Courty03a},
\begin{equation}
-i\Omega Z_{\rm m}X_{\rm m}=\hbar \xi _{\rm a}a_0^{\rm in}+F_{\rm
m}, \label{Equ_ZmVm}
\end{equation}
where $Z_{\rm m}$ is the mechanical impedance of the mirror. The first force
is the radiation pressure of the intracavity field, expressed in terms of the
incident intensity fluctuations $a_0^{\rm in}$. The second force $F_{\rm m}$
represents the classical coupling with the environment.

The information provided by the phase of the reflected field is described by
an estimator $\hat{X}_{\rm sig}$ of the measurement which is obtained through
a normalization of the output phase (\ref{Equ_Phase}) as a displacement. It
appears as the sum of the signal and extra noise terms,
\begin{equation}
\hat{X}_{\rm sig}=\frac{1}{2\xi _{\rm a}}a_{\frac{\pi }{2}}^{\rm out}=X_{\rm
sig}+\frac{1}{2\xi _{\rm a}}a_{\frac{\pi }{2}}^{\rm in}+X_{\rm m}.
\label{Equ_Xsig}
\end{equation}
The sensitivity is limited by an equivalent input noise equal to the noises
added in the estimator. Since all these noises are uncorrelated for an
incident coherent light, the equivalent input noise spectrum $\Sigma _{\rm
sig}$ is given by,
\begin{eqnarray}
\Sigma _{\rm sig}&=&\frac{1}{4\xi _{\rm a}^2}+\sigma _{X_{\rm
m}X_{\rm m}}, \label{Equ_Ssig} \\
&=&\frac{1}{4\xi _{\rm a}^2}+\frac{\hbar ^2\xi _{\rm a}^2}{\Omega
^2\left| Z_{\rm m}\right| ^2}+\frac{\sigma _{F_{\rm m}F_{\rm
m}}}{\Omega ^2\left| Z_{\rm m}\right| ^2}, \label{Equ_Ssigfree}
\end{eqnarray}
where $\sigma _{X_{\rm m}X_{\rm m}}$ and $\sigma _{F_{\rm m}F_{\rm m}}$ are
the noise spectra of the displacement $X_{\rm m}$ and of the classical force
$F_{\rm m}$. The first term in (\ref{Equ_Ssigfree}) is the measurement error
due to the incident phase noise, the second term the back-action noise due to
radiation pressure, and the last one the classical noise. Curve {\it a} in
figure \ref{Fig_Sensit} shows the quantum-limited sensitivity obtained by
neglecting this last term, and considering a suspended mirror for which the
impedance reduces to
\begin{equation}
Z_{\rm m}\simeq -i\Omega M_{\rm m}, \label{Equ_Zm}
\end{equation}
where $M_{\rm m}$ is the mirror mass. Radiation pressure is dominant at low
frequency with a $1/\Omega ^4$ dependence of the noise power spectrum,
whereas phase noise is dominant at high frequency with a flat frequency
dependence, at least for frequencies smaller than the cavity bandwidth. Curve
{\it b} is the so-called standard quantum limit which corresponds to the
minimum noise reachable at a given frequency by varying the optomechanical
coupling $\xi _{\rm a}$. For a given value of $\xi _{\rm a}$, the sensitivity
is optimal at only one frequency defined as,
\begin{equation}
\Omega _{\rm a}^{\rm SQL}=\sqrt{2\hbar \xi _{\rm a}^2/M_{\rm m}}.
\label{Equ_OSQL}
\end{equation}

Squeezed states may change this behavior. Equations (\ref{Equ_ZmVm}) and
(\ref{Equ_Xsig}) show that the input noise is related to a specific
combination of the incident intensity and phase quadratures. For a suspended
mirror it is proportional to a particular incident quadrature $a_{-\theta
}^{\rm in}$, with an angle $\theta $ defined by,
\begin{eqnarray}
\hat{X}_{\rm sig}&=&X_{\rm sig}-\frac{1}{2\xi _{\rm a}\sin \theta }a_{-\theta
}^{\rm in}, \label{Equ_XsigSqueez} \\
\cot \theta &=&\left( \Omega _{{\rm a}}^{{\rm SQL}}/\Omega \right) ^{2}.
\label{Equ_ThetaOpt}
\end{eqnarray}
The sensitivity of the measurement is then improved by using an incident
squeezed state for which the noise of this quadrature is reduced
\cite{Jaekel90}. Note that the optimal angle $\theta $ is frequency dependent
so that the squeezing angle must vary with frequency. This can be done by
sending a squeezed state with a constant squeezing angle in a detuned cavity
\cite{Kimble02}.

Another possibility to improve the sensitivity is to take advantage of the
self phase-modulation induced by radiation pressure to perform a back-action
evading measurement. Instead of detecting the output phase quadrature, we
measure an other quadrature $a_\theta ^{\rm out}$. According to equations
(\ref{Equ_Ampl}) and (\ref{Equ_Phase}), the input-output relation for this
quadrature is given by,
\begin{equation}
\frac{1}{2\xi _{\rm a}\sin \theta }a_\theta ^{\rm out}=\frac{1}{2\xi _{\rm
a}}a_\frac{\pi }{2}^{\rm in}+X_{\rm sig}+X_{\rm m}+\frac{\cot \theta }{2\xi
_{\rm a}}a_0^{\rm in}. \label{Equ_ioQuad}
\end{equation}
In the case of a suspended mirror, the two last terms cancel out for the
particular quadrature angle given by (\ref{Equ_ThetaOpt}). Radiation-pressure
effects associated with the motion of mirror $m$ then disappear in the
measurement and the sensitivity is only limited by the phase noise [first
term in (\ref{Equ_ioQuad})]. As in the case of the injection of squeezed
state, beating the standard quantum limit over a wide bandwidth requires to
adapt the detected quadrature to the frequency dependence of
radiation-pressure effects \cite{Kimble02}. This can be done by sending the
reflected field in properly optimized detuned cavities.

In both cases, the sensitivity improvement strongly depends on the quantum
properties of the optomechanical coupling between the light and the suspended
mirror. It is in particular necessary to avoid losses in the interferometric
measurement.

\section{Quantum locking of a mirror}
\label{Sec_QuantumLocking}

\begin{figure}
\begin{center}
\includegraphics[width=7.7cm]{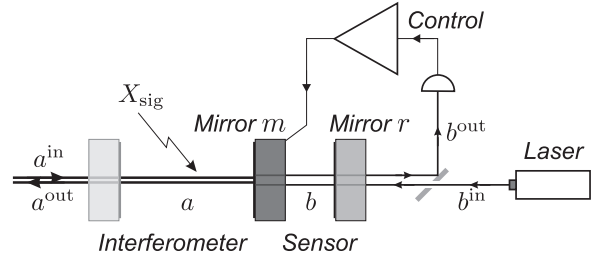}
\end{center}
\caption {\label{Fig_Scheme}
Active control of mirror $m$. Its motion is monitored by a second
cavity based on a reference mirror $r$, and controlled by a
feedback loop.}
\end{figure}

Figure \ref{Fig_Scheme} shows the scheme used to perform a quantum
locking of mirror $m$. The mirror motion is monitored by another
cavity made of the mirror $m$ itself and a reference mirror $r$.
The information is fed back to the mirror in order to control its
displacements. The field $b$ in this control cavity obeys
equations similar to the ones of field $a$, except for the cavity
length variation now equal to the relative displacement $X_{\rm
r}-X_{\rm m}$ between the two mirrors. According to equations
(\ref{Equ_Ampl}) and (\ref{Equ_Phase}), the phase of the reflected
field provides an estimator $\hat{X}_{\rm m}$ for the motion of
mirror $m$, with a sensitivity limited by the incident phase noise
and the motion of mirror $r$,
\begin{equation}
\hat{X}_{\rm m}=-\frac{1}{2\xi _{\rm b}}b_{\frac{\pi }{2}}^{\rm out}=X_{\rm
m}-\frac{1}{2\xi _{\rm b}}b_{\frac{\pi }{2}}^{\rm in}-X_{\rm r},
\label{Equ_Xmest}
\end{equation}
where $\xi _{\rm b}$ is the optomechanical parameter for the control cavity.

The mirror $m$ is submitted to a feedback force proportional to
this estimator. One has also to take into account the radiation
pressures from both cavities, so that the motions of mirrors $m$
and $r$ are given by,
\begin{eqnarray}
-i\Omega Z_{\rm m}X_{\rm m}&=&\hbar \xi _{\rm a}a_0^{\rm in}+F_{\rm m}-\hbar
\xi _{\rm b}b_0^{\rm in}+i\Omega Z_{\rm fb}\hat{X}_{\rm m},
\label{Equ_ZmVm2}\\
-i\Omega Z_{\rm r}X_{\rm r}&=&\hbar \xi _{\rm b}b_0^{\rm in}+F_{\rm r},
\label{Equ_ZrVr}
\end{eqnarray}
where $F_{\rm r}$ represents the classical noise of mirror $r$ and $Z_{\rm
fb}$ is the transfer function of the feedback loop. The resulting motion of
mirror $m$ is obtained from equations (\ref{Equ_Xmest}) and
(\ref{Equ_ZmVm2}),
\begin{equation}
-i\Omega \left( Z_{\rm m}+Z_{\rm fb}\right) X_{\rm m}=\hbar \xi _{\rm
a}a_0^{\rm in}+F_{\rm m}-\hbar \xi _{\rm b}b_0^{\rm in}-i\Omega Z_{\rm
fb}\left( X_{\rm r}+\frac{1}{2\xi _{\rm b}}b_{\frac{\pi }{2}}^{\rm in}\right)
. \label{Equ_Vmfb}
\end{equation}
The control changes the response of mirror $m$ by adding a feedback-induced
impedance $Z_{\rm fb}$ to the mechanical impedance $Z_{\rm m}$. For a large
feedback gain, the effective impedance is increased and the mirror
displacements are reduced. The control also contaminates the mirror motion by
the noises in the control cavity [last terms in (\ref{Equ_Vmfb})].

We first examine the effect of the control on classical noise,
neglecting all the quantum noises. According to equations
(\ref{Equ_ZrVr}) and (\ref{Equ_Vmfb}), the resulting motion of
mirror $m$ is given by,
\begin{equation}
-i\Omega Z_{\rm m}X_{\rm m}\simeq \frac{Z_{\rm m}F_{\rm m}+Z_{\rm
fb}F_{\rm r}}{Z_{\rm m}+Z_{\rm fb}},
\end{equation}
where we have assumed for simplicity the two mirrors identical so
that $Z_{{\rm r}}=Z_{{\rm m}}$. As the feedback gain increases,
the classical force $F_{\rm m}$ applied on mirror $m$ is replaced
by the force $F_{\rm r}$ acting on mirror $r$, which can be less
noisy if the reference mirror is less coupled to its environment.
From the expression (\ref{Equ_Ssig}) of the equivalent input noise
$\Sigma _{\rm sig}$, the classical noise in the interferometric
measurement is reduced down to the displacement noise of mirror
$r$. In other words, the control locks the motion of mirror $m$ to
the one of the reference mirror $r$, leading to a transfer of
noise from the sensor measurement to the interferometric one.

\begin{figure}
\begin{center}
\includegraphics[height=4.4cm]{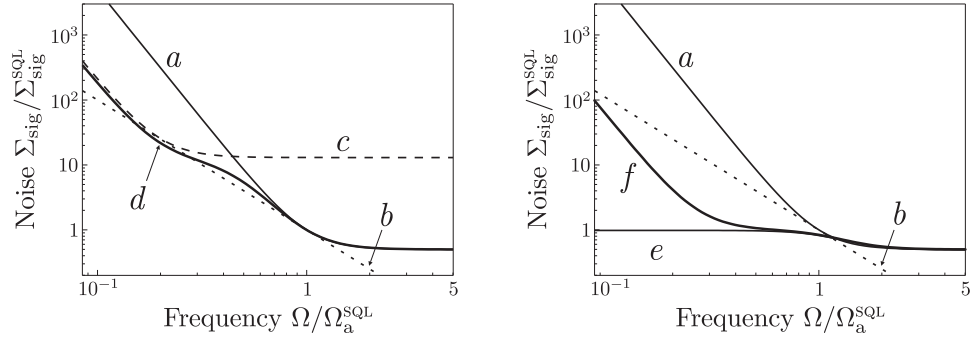}
\end{center}
\caption {\label{Fig_Sensit} Equivalent input noise $\Sigma _{\rm sig}$ in
the interferometric measurement as a function of frequency $\Omega $: free
interferometer ({\it a}), standard quantum limit ({\it b}), quantum locking
with $\xi _{\rm b}=\xi _{\rm a}/5$ for infinite ({\it c}) and optimum ({\it
d}) gains, back-action cancellation with $\xi _{\rm b}=\xi _{\rm a}$ ({\it
e}), and with 1\% loss in the control cavity ({\it f}). Noise is normalized
to $1/2\xi _{\rm a}^2$ and frequency to $\Omega _{\rm a}^{\rm SQL}$.}
\end{figure}

A similar transfer of sensitivity occurs at the quantum level. For a very
large feedback gain ($Z_{\rm fb}\rightarrow \infty $), only the last term in
(\ref{Equ_Vmfb}) is significant, and one gets
\begin{equation}
X_{\rm m}\simeq X_{\rm r}+\frac{1}{2\xi _{\rm b}}b_{\frac{\pi
}{2}}^{\rm in}.
\end{equation}
The motion of mirror $m$ no longer depends on the radiation pressure in the
interferometric measurement. Apart from the phase noise of beam $b$, the
mirror $m$ is locked at the quantum level on the reference mirror $r$. Its
displacement noise reproduces the quantum noises in the sensor measurement,
and the resulting sensitivity for the interferometric measurement, deduced
from (\ref{Equ_Ssig}) and (\ref{Equ_ZrVr}), is given by
\begin{equation}
\Sigma _{\rm sig}^\infty =\frac{1}{4\xi _{\rm a}^2}+\frac{1}{4\xi
_{\rm b}^2}+\frac{\hbar ^2\xi _{\rm b}^2}{\Omega ^2\left| Z_{\rm
r}\right| ^{2}}. \label{Equ_Ssiginf}
\end{equation}
The two last terms in this equation correspond to the equivalent
input noise for the sensor measurement [compare to equation
(\ref{Equ_Ssigfree}) with $\xi _{{\rm a}}$ replaced by $\xi _{{\rm
b}}$, and $Z_{{\rm m}}$ by $Z_{{\rm r}}$]. This quantum transfer
of noises is shown in curve {\it c} of figure \ref{Fig_Sensit},
obtained with an optomechanical parameter $\xi _{{\rm b}}$ equal
to $\xi _{{\rm a}}/5$. Since the sensor measurement is less
sensitive than the interferometric one ($\xi _{{\rm b}}<\xi _{{\rm
a}}$), radiation-pressure effects of beam $b$ are smaller than the
ones of beam $a$. At low frequency where these effects are
dominant, the mirror $m$ reproduces the motion of the reference
mirror, leading to a clear reduction of noise.

At high frequency, the sensitivity is contaminated by the phase noise in the
sensor measurement. This can easily be improved by using a
frequency-dependent feedback gain in such a way that the control is efficient
at low frequency whereas it plays no significant role at high frequency.
Curve {\it d} of figure \ref{Fig_Sensit} shows the result obtained by an
optimization of the feedback gain at every frequency. One gets a very clear
noise reduction at low frequency while the sensitivity at high frequency is
preserved.

\section{Back-action cancellation}
\label{Sec_BackAction}

The quantum locking presented in the previous section cancels the
radiation-pressure effects in the interferometer, replacing them by the less
noisy effects in the sensor measurement. It is possible to completely
suppress the back-action noise in the sensor measurement by using the quantum
optics techniques presented in section \ref{Sec_QuantumLimits}, such as the
injection of squeezed states \cite{Jaekel90} or the optimization of the
detected quadrature \cite{Kimble02}. Using these techniques in the sensor
measurement rather than in the interferometer itself presents the advantage
that all the necessary adaptations has to be made on the sensor and not on
the interferometer.

As in the case of the optimization of the detected quadrature for the free
interferometer [equation (\ref{Equ_ioQuad})], measuring a quadrature
$b_{\theta }^{\rm out}$ different from the phase quadrature changes the
estimator $\hat{X}_{\rm m}$ by adding a term proportional to the incident
intensity fluctuations $b_{0}^{\rm in}$,
\begin{equation}
\hat{X}_{\rm m}=-\frac{1}{2\xi _{\rm b}\sin \theta }b_\theta ^{\rm
out}=X_{\rm m}-\frac{1}{2\xi _{\rm b}}b_{\frac{\pi }{2}}^{\rm in}-X_{\rm
r}-\frac{\cot \theta }{2 \xi_ {\rm b}}b_{0}^{\rm in}. \label{Equ_Xmest2}
\end{equation}
Different optimizations of the detected quadrature are possible. Since both
mirror motions $X_{\rm m}$ and $X_{\rm r}$ depend on radiation pressure in
the sensor cavity, one can eliminate the whole contribution or only the
contribution due to the reference mirror. Considering two identical and
suspended mirrors, the second solution is simpler since it corresponds to an
angle $\theta $ given by an equation similar to (\ref{Equ_ThetaOpt}) with
$\Omega _{\rm a}^{\rm SQL}$ replaced by the frequency $\Omega _{\rm b}^{\rm
SQL}$ defined as in (\ref{Equ_OSQL}). This angle is experimentally accessible
by sending the field in a single detuned cavity \cite{Courty03b}. It
furthermore corresponds to a back-action evading measurement of the motion of
mirror $m$ by the sensor cavity. Neglecting the classical noise in the motion
(\ref{Equ_ZrVr}) of the reference mirror, the estimator $\hat{X}_{\rm m}$ of
the sensor measurement indeed reduces for this value of $\theta $ to,
\begin{equation}
\hat{X}_{\rm m}=X_{\rm m}-\frac{1}{2\xi _{\rm b}}b_{\frac{\pi }{2}}^{\rm in}.
\label{Equ_Xmest3}
\end{equation}
As compared to the standard detection scheme [equation (\ref{Equ_Xmest})],
the measurement is no longer sensitive to radiation-pressure effects in the
sensor. Its sensitivity is only limited by the incident phase noise which can
be made arbitrarily small by increasing the optomechanical coupling $\xi
_{\rm b}$. For an infinite feedback gain, the control freezes the motion of
mirror $m$ down to a limit associated with the phase noise,
\begin{equation}
X_{\rm m}\simeq \frac{1}{2\xi _{\rm b}}b_{\frac{\pi }{2}}^{\rm in},
\end{equation}
and the resulting sensitivity for the interferometric measurement is deduced
from (\ref{Equ_Ssig}),
\begin{equation}
\Sigma _{\rm sig}^\infty =\frac{1}{4\xi _{\rm a}^2}+\frac{1}{4\xi _{\rm
b}^2}. \label{Equ_Ssiginf2}
\end{equation}
The sensitivity reduces to the sum of phase noises of both cavities. Curve
{\it e} of figure \ref{Fig_Sensit} shows the sensitivity obtained with an
optimized feedback gain. Radiation-pressure effects are completely
suppressed, resulting in a sensitivity only limited by the phase noises and
almost flat over the whole frequency band.

An essential feature of the active control is to be decoupled from other
optimizations of the interferometer. Losses in the interferometer usually
have a drastic effect on the noise reduction obtained by quantum optics
techniques \cite{Kimble02}. Quantum locking, however, is insensitive to such
losses and they have no effect on the sensitivity improvement obtained with
this technique. We have actually made no assumption on the exact motion of
mirror $m$. The back-action cancellation in the sensor measurement and the
quantum locking of the mirror do not depend on the optomechanical coupling
between the mirror and the light in the interferometer. Imperfections in the
interferometer thus do not affect the control.

As usual in quantum optics, losses must be avoided in the optomechanical
sensor. Curve {\it f} in figure \ref{Fig_Sensit} shows the sensitivity
obtained with 1\% loss in the control cavity \cite{Courty03b}. In contrast to
the lossless case (curve {\it e}), back-action cancellation is no longer
perfect. One however still has a very large reduction of radiation-pressure
noise as compared to the free interferometer.

\section{Quantum locking of a cavity}
\label{Sec_CavityLocking}

\begin{figure}
\begin{center}
\includegraphics[width=7.7cm]{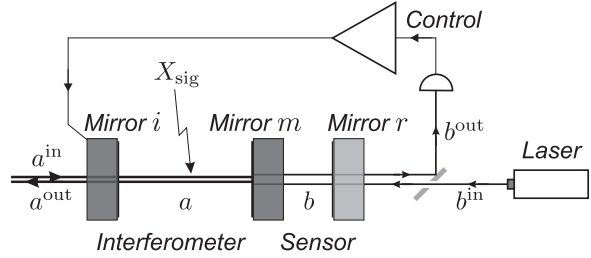}
\end{center}
\caption {\label{Fig_Scheme2} Active control of the whole interferometric
cavity by a single feedback mechanism. The motions of mirrors $m$ and $r$ are
measured by the sensor cavity. The information is fed back to the input
mirror $i$ in order to suppress the length variations of the interferometric
cavity induced by radiation pressure.}
\end{figure}

The input mirror of the interferometric cavity (mirror $i$ in figure
\ref{Fig_Scheme2}) moves as well in response to radiation pressure and a
complete control of the interferometric measurement would require a local
control of each mirror of the cavity. We show in this section that a single
control mechanism can lock the cavity at the quantum level \cite{Giazotto}.

Taking into account the motion of mirror $i$, the estimator of the
interferometric measurement now depends on the differential motion
between mirrors $m$ and $i$, and equation (\ref{Equ_Xsig}) is
modified to
\begin{equation}
\hat{X}_{\rm sig}=X_{\rm sig}+\frac{1}{2\xi _{\rm a}}a_{\frac{\pi
}{2}}^{\rm in}+X_{\rm m}-X_{\rm i}. \label{Equ_Xsig2}
\end{equation}

As shown in figure \ref{Fig_Scheme2}, the principle of the locking
is to use the information provided by the sensor measurement to
apply a feedforward force to the input mirror $i$, in such a way
that its motion follows the one of mirror $m$. The key point is
that the sensor gives access to the differential motion $X_{\rm
m}-X_{\rm i}$ between mirrors $m$ and $i$, for an appropriate
choice of the detected quadrature. Neglecting the classical noise,
the motion of mirror $i$ without feedback is indeed related to the
radiation pressure in the interferometric cavity,
\begin{equation}
-i\Omega Z_{\rm i}X_{\rm i}=-\hbar \xi _{\rm a}a_0^{\rm in}.
\label{Equ_ZiVi}
\end{equation}
Equations (\ref{Equ_ZmVm2}), (\ref{Equ_ZrVr}) and (\ref{Equ_ZiVi}) show that
the total radiation-pressure force exerted on the system composed of the
three mirrors is equal to zero. The motions of the three mirrors are not
independent and one gets,
\begin{equation}
Z_{\rm i}X_{\rm i}+Z_{\rm m}X_{\rm m}+Z_{\rm r}X_{\rm r}=0.
\label{Equ_XiXmXr}
\end{equation}
In the case of identical and suspended mirrors, the sum of the
three displacements cancels out. We then detect a quadrature
$b_{\theta }^{\rm out}$ at the output of the sensor cavity with an
angle $\theta $ defined by
\begin{equation}
\cot \theta =\frac{3}{2}\left( \Omega _{{\rm b}}^{{\rm
SQL}}/\Omega \right) ^{2}. \label{Equ_ThetaOpt2}
\end{equation}
As in the previous section, this quadrature is obtained by sending
the field $b^{\rm out}$ in a properly detuned cavity
\cite{Courty03b}. According to (\ref{Equ_Xmest2}), the estimator
of the sensor measurement is given by
\begin{equation}
\hat{X}_{\rm m}=X_{\rm m}+\frac{1}{2}X_{\rm r}-\frac{1}{2\xi _{\rm
b}}b_{\frac{\pi }{2}}^{\rm in}=\frac{1}{2}\left( X_{\rm m}-X_{\rm
i}\right) -\frac{1}{2\xi _{\rm b}}b_{\frac{\pi }{2}}^{\rm in},
\label{Equ_Xmest4}
\end{equation}
where we have used the relation $X_{\rm r}=-X_{\rm m}-X_{\rm i}$.
Apart from the phase noise of beam $b$, the sensor measures the
differential motion $X_{\rm m}-X_{\rm i}$.

The quantum locking is obtained by applying a feedforward force
$-i\Omega Z_{\rm fb}\hat{X}_{\rm m}$ to mirror $i$, with a
feedforward gain $Z_{\rm fb}$ equal to 2. This force induces a
displacement of mirror $i$ by a quantity $2\hat{X}_{\rm m}$, and
its resulting motion is given by,
\begin{equation}
X_{\rm i}\rightarrow X_{\rm i}+2\hat{X}_{\rm m}=X_{\rm
m}-\frac{1}{\xi _{\rm b}}b_{\frac{\pi }{2}}^{\rm in}.
\label{Equ_Xi}
\end{equation}
The mirror $i$ is then locked on the mirror $m$. The differential
motion $X_{\rm m}-X_{\rm i}$ no longer depends on radiation
pressure and the sensitivity of the measurement, deduced from
(\ref{Equ_Xsig2}), reduces to the phase noises,
\begin{equation}
\Sigma _{\rm sig} =\frac{1}{4\xi _{\rm a}^2}+\frac{1}{\xi _{\rm
b}^2}. \label{Equ_Ssig2}
\end{equation}
For an optomechanical parameter $\xi_{\rm b}$ larger than
$\xi_{\rm a}$, the sensitivity is only limited by the phase noise
$1/4\xi _{\rm a}^2$ in the interferometer. Radiation-pressure
effects are completely suppressed.

Although this quantum locking is not a local control as in the
previous sections, an important feature is that it is still
insensitive to losses in the interferometer. As a matter of fact,
we have made no assumption on the radiation pressure in the
interferometer, except that both mirrors $i$ and $m$ are submitted
to the same force [equations (\ref{Equ_ZmVm2}) and
(\ref{Equ_ZiVi})]. Losses in a gravitational-wave interferometer
are mainly due to imperfections on the mirrors, but the
propagation between the two mirrors usually is lossless. Radiation
pressures exerted on each mirror are the sum of the radiation
pressures of the incoming and outgoing fields, and they are the
same on both mirrors whatever the mirror losses are. Relation
(\ref{Equ_XiXmXr}) and the principle of the quantum locking are
then valid even in presence of losses in the interferometer.

\section{Conclusion}

The active control studied in this paper is based on a local optomechanical
sensor which measures the position fluctuations of the mirror and locks its
position with respect to a reference mirror. The main characteristic of the
sensor is its sensitivity, defined by an optomechanical parameter $\xi _{{\rm
b}}$ which depends on the cavity finesse and the light power. Quantum locking
is efficient when the optomechanical parameter of the sensor is of the same
order as the one of the interferometer. This seems easy to achieve with
currently available technology \cite{Rempe92,Cohadon99}. Taking for example
the parameters of the V{\sc irgo} interferometer ($15$ $kW$ light power in
each Fabry-Perot arms with a global finesse of $600$ \cite{Bradaschia90}),
this condition corresponds to a sensor of finesse $10^{5}$ with an
intracavity light power of $90$ $W$, that is an incident light power of $1.5$
$mW$ only. Due to the high finesse of the cavity, the same sensitivity is
reached for the sensor than for the interferometer itself, while the
intracavity and incident light powers are much smaller.

This technique is useful to reduce classical noise such as thermal
fluctuations, as long as the reference mirror of the sensor cavity
is less noisy than the mirror of the interferometer. This has
already been experimentally demonstrated for the thermal noise of
internal acoustic modes of a mirror \cite{Cohadon99}. In this
case, only the cooled mirror is resonant at frequency of interest
and thermal noise reduction as large as 1000 have been obtained
\cite{Pinard00}. This technique may be of some help for cryogenic
gravitational-wave interferometers. A major issue is the heat
generation due to the absorption of the high-power light in the
interferometer, which prevents from cooling to very low
temperatures \cite{Uchiyama98}. Since the light in the sensor is
much less intense, a low temperature can be reached for the
reference mirror by passive cryogenic cooling, and transferred to
the interferometer mirror by active control. Concerning the
internal thermal noise of mirrors, however, the sensor cavity must
detect the noise as it is seen by the interferometric measurement.
The optical waist in the sensor cavity must be adapted to the one
in the interferometer, requiring to develop high-finesse cavities
with large effective waists \cite{Marin03}.

We have shown that a local control of mirrors allows one to
efficiently reduce the quantum effects of radiation pressure in an
interferometric measurement. The back-action noise is completely
suppressed by using an optimized detection strategy. The
sensitivity is thus greatly improved in the low-frequency domain
where radiation-pressure effects are dominant, without alteration
in the high-frequency domain where phase noise prevails.

In a practical implementation, the complete control of a gravitational-wave
interferometer would require the use of optomechanical sensors for each
sensitive mirrors, that is the four mirrors of the Fabry-Perot cavities in
the interferometer arms. We have shown that a single control mechanism can
lock the whole cavity at the quantum level. Adjusting the detected
quadrature, an optomechanical sensor placed near the end mirror of the cavity
can monitor the cavity length variations induced by radiation pressure. The
information is then fed back to the front mirror in order to suppress
radiation-pressure effects in the interferometric measurement.

Finally note that it is also possible to perform a correction of
the signal delivered by the interferometer rather than to control
the mirror motion. In that case, the optomechanical sensor detects
the radiation-pressure effects in the interferometer, and the
information is numerically subtracted from the result of the
interferometric measurement. As long as the effect of the mirror
motions on the output of the interferometer is known with a
sufficient accuracy, the two techniques are in principle
equivalent.

These results show that active control is a powerful technique to reduce
quantum noise. An essential characteristic of this approach is to be
decoupled from other optimizations of the interferometer. As usual in quantum
optics, losses must be avoided in the optomechanical sensor. Quantum locking,
however, is insensitive to imperfections in the interferometer. All the
necessary adaptations has to be made on the control measurement, but the
quantum locking does not induce any additional constraint on the
interferometer.

\ack

We thank Adalberto Giazotto and Giancarlo Cella for fruitful
discussions.

\end{document}